\documentclass[twocolumn]{aastex62}
\usepackage{amsmath,amstext}
\usepackage{comment}
\usepackage{apjfonts} 
\usepackage{chngcntr}

\newcommand{\be}{\begin{eqnarray}}
\newcommand{\ee}{\end{eqnarray}}
\newcommand{\lp}{\left(}
\newcommand{\rp}{\right)}

\shorttitle{Super-puffs as Ringed Exoplanets}
\shortauthors{Piro \& Vissapragada}

\begin{document}

\title{\Large \textbf{Exploring Whether Super-Puffs Can Be Explained as Ringed Exoplanets}}

\author{Anthony L. Piro}
\affiliation{The Observatories of the Carnegie Institution for Science, 813 Santa Barbara St., Pasadena, CA 91101, USA; piro@carnegiescience.edu}

\author{Shreyas Vissapragada}
\affiliation{Division of Geological and Planetary Sciences, California Institute of Technology, 1200 E California Blvd., Pasadena, CA 91125, USA; svissapr@caltech.edu}

\begin{abstract}
An intriguing, growing class of planets are the ``super-puffs,'' objects with exceptionally large radii for their masses and thus correspondingly low densities ($\lesssim0.3\rm\,g\,cm^{-3}$). {Here we consider whether they could have large inferred radii because they are in fact ringed.} This would naturally explain why super-puffs have thus far only shown featureless transit spectra. We find that this hypothesis can work in some cases but not all. The close proximity of the super-puffs to their parent stars necessitates rings with a rocky rather than icy composition. This limits the radius of the rings, and makes it challenging to explain the large size of Kepler~51b, 51c, 51d, and 79d unless the rings are composed of porous material. Furthermore, the short tidal locking timescales for Kepler~18d, 223d, and 223e mean that these planets may be spinning too slowly, resulting in a small oblateness and rings that are warped by their parent star. Kepler~87c and 177c have the best chance of being explained by rings. {Using transit simulations, we show that testing this hypothesis requires photometry with a precision of somewhere between $\sim10$~ppm and $\sim50$~ppm, which roughly scales with the  ratio of the planet and star's radii.} We conclude with a note about the recently discovered super-puff HIP~41378f.
\end{abstract}

\keywords{
        occultations ---
        planets and satellites: detection ---
        planets and satellites: rings }

\section{Introduction}

In our solar system, rings are common amongst the four outer gas-rich planets as well as present for some of the smaller rocky bodies. Nevertheless, discovering rings around planets outside our solar systems has been challenging. In principle, rings should be detectable from detailed photometric or spectroscopic changes to transits \citep[e.g.,][]{Barnes04,Ohta09,Santos15,Zuluaga15,Akinsanmi18}. The difficulty is that such signals are subtle and difficult to discern in current data. In a few cases, potential rings or constraints on rings have been made in this way \citep{Heising15,Aizawa17,Aizawa18}, and in at least one instance it has been argued that an exoplanet has a giant ring system from a series of complex eclipses \citep{Kenworthy15,Rieder16}. There is clearly still a lot we do not know about the rings of exoplanets.

The simplest impact of rings is to increase the depth of transits so that instead of measuring the planet radius $R_p$, an eclipsed area of $A$ results in an inferred radius of \citep{Piro18b}
\be
	R_{\rm inf} = (A/\pi)^{1/2}\gtrsim R_p.
    \label{eq:r_inf}
\ee
A useful example to consider is that of Saturn: averaged over season, if an external observer measured Saturn's size in transit without accounting for rings, they would underestimate its true density by about a factor of two. Thus if a population of exoplanets are found with anomalously large radii, and correspondingly low densities, this may indicate we are observing $R_{\rm inf}$ rather than $R_p$.

In fact, there is a growing class of exoplanets with inferred densities of $\lesssim 0.3\,{\rm g\,cm^{-3}}$, also known as ``super-puffs'' \citep{Cochran11,Masuda14,Jontof-Hutter14,Ofir14,Mills16,Vissapragada19,Santerne19}. The properties of these planets and their host stars are summarized in Table \ref{table}. We note that different authors have differing definitions for this class of planets, with some adopting a strict boundary of $M_p < 10M_\Earth$ \citep[e.g.,][]{Lee19,Jontof-Hutter19}. Here we take the slightly more liberal approach of including planets with $M_p \lesssim 15\,M_\Earth$, which includes the low-density planets Kepler-18d \citep{Cochran11} and Kepler-177c \citep{Vissapragada19}.

This new class of planets with larger radii than expected bears some similarity to the classical problem of hot Jupiter radius inflation. However, hot Jupiter inflation is strongly correlated with equilibrium temperature \citep{Miller11,Thorngren18}, which means that a similar mechanism cannot be extended to the much cooler super-puffs. While some super-puff systems are young, and therefore may appear inflated because they are still contracting \citep{LibbyRoberts19}, most of these planets are older and cannot be explained with youth either. Other proposed explanations for these planets include dusty outflows \citep{Wang19}, photochemical hazes \citep{Kawashima19}, inflation from tidal heating \citep{Millholland19}, or especially thick gas envelopes \citep{Lee16}. If it is the latter, then these exoplanets would be prime targets for transit spectroscopy, but when this has been performed the results are featureless spectra \citep{LibbyRoberts19}.

Here we consider the alternative hypothesis that super-puffs are in fact ringed exoplanets. In Section~\ref{sec:basic}, we explore whether super-puffs can be explained as planets with rings, and what this implies about both the rings and the underlying planets. We find that this explanation works for some of the super-puffs, but for others it has difficulties. In Section~\ref{sec:detectability}, we perform transit simulations to assess whether this hypothesis can be constrained by current or future observational efforts. We then conclude in Section~\ref{sec:conclusions} with a summary of this work.

\section{Constraints from the Ring Hypothesis}
\label{sec:basic}

  \begin{deluxetable*}{lcccccccccc}
  \tablecolumns{11} \tablewidth{900pt}
 \tablecaption{Super-puff Planet and Parent Star Properties}
   \tablehead{Name & $M_p$ ($M_\Earth$) & $R_{\rm inf}$ ($R_\Earth$) & $\langle\rho\rangle$ (g\,cm$^{-3}$) & $a$ (AU)& $P$ (days) & $M_*$ ($M_\odot$) & $R_*$ ($R_\odot$) & $T_{\rm eff}$ (K) & $(R_{\rm inf}/R_*)^2$ & Reference}
  \startdata
    Kepler~18d & 16.4 & 6.98 & 0.27  & 0.12  & 14.86 & 0.97 & 1.11 & 5345 & $3.3\times10^{-3}$ & (1)\\
    Kepler~51b & 3.7 & 6.89 & 0.06 & 0.25 & 45.15 & 0.99 & 0.88 & 5670 & $5.1\times10^{-3}$ & (2,3)\\
	Kepler~51c & 4.4 & 8.98 & 0.03 & 0.38 & 85.31 & $\cdots$ & $\cdots$ & $\cdots$ & $8.7\times10^{-3}$ & $\cdots$ \\
    Kepler~51d & 5.7 & 9.46 & 0.04 & 0.51 & 130.18 & $\cdots$ & $\cdots$ & $\cdots$ & $9.7\times10^{-3}$ & $\cdots$\\
    Kepler~79d & 6.0 & 7.16 & 0.09 & 0.29 & 52.09 & 1.17 & 1.30 & 6174 & $2.5\times10^{-3}$ & (4)
    \\
    Kepler~87c & 6.4 & 6.14 & 0.15 & 0.68 & 191.23 & 1.10 & 1.82 & 5600 & $9.6\times10^{-4}$ & (5)\\
    Kepler~177c & 14.7 & 8.73 & 0.12 & 0.26 & 49.41 & 0.92 & 1.32 & 5732 & $3.6\times10^{-3}$ & (6)\\
    Kepler~223d & 8.0 & 5.24 & 0.31 & 0.13 & 14.79 & 1.13 & 1.72 & 5821 & $7.8\times10^{-4}$ & (7)\\
    Kepler~223e & 4.8 & 4.60 & 0.28 & 0.15 & 19.73 & $\cdots$ & $\cdots$ & $\cdots$  & $6.0\times10^{-4}$ & $\cdots$\\
    HIP~41378f & 12 & 9.2 & 0.09 & 1.37 & 542.08 & 1.16 & 1.27 & 6320 & $4.4\times10^{-3}$ & (8)\\
   \enddata
\tablecomments{(1)~\citet{Cochran11}, (2)~\citet{Masuda14}, (3)~\citet{LibbyRoberts19}, (4)~\citet{Jontof-Hutter14}, (5)~\citet{Ofir14}, (6)~\citet{Vissapragada19}, (7)~\citet{Mills16},
(8)~\citet{Santerne19}}
\end{deluxetable*}
  \label{table}

We first consider the hypothesis that super-puffs are actually planets with rings, and investigate what this implies about the properties of such rings and the planets themselves.
  
\subsection{Constraints on Ring Material}

The rings of Saturn would be the closest analog to what we are considering here, since super-puff rings must be extended and optically thick if they are to cause such large inferred radii. An important difference in comparison to Saturn is that super-puffs are much closer to their parent stars. Ice sublimates at a temperature $T_{\rm sub}\approx170\,{\rm K}$, so rings cannot be composed of ice for planets with a semi-major axis within \citep{Gaudi03}
\be
    a \lesssim \lp \frac{L_*}{16\pi \sigma_{\rm SB}T_{\rm sub}^4} \rp^{1/2}
    \approx 2.7 \lp\frac{L_*}{L_\odot} \rp^{1/2} {\rm AU}.
    \label{eq:sublimate}
\ee
For this reason, the material forming the rings of super-puffs will be rocky with a typical density in the range of \mbox{$\approx2-5\,{\rm g\,cm^{-3}}$,} depending on the exact composition. This is only approximate, and may extend somewhat below $2\,{\rm g\,cm^{-3}}$ for material with a higher porosity.

\begin{figure}
\epsscale{1.0}
  \includegraphics[width=0.48\textwidth,trim=1.0cm 0.0cm 0.0cm 0.0cm]{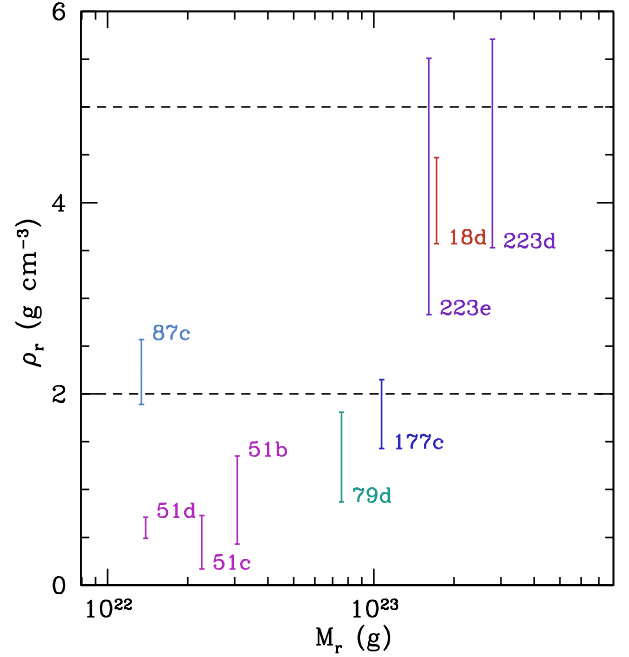}
\caption{Ring material density $\rho_r$ and mass $M_r$ constraints if super-puffs were instead ringed using Equations (\ref{eq:rhor}) and (\ref{eq:mr}), respectively. Super-puffs that lie between the two horizontal dashed lines can potentially have their large radii explained as rocky rings. Those below the lower dashed line have difficulty being explained in this way unless the ring material is more porous. The ring mass is required for the ring to last a timescale of $t_{\rm PR}=10^9\,{\rm yr}$ with an inclination angle of $i=45^\circ$. The range of ring masses needed is comparable to large asteroids in our solar system.}
\label{fig:composition}
\epsscale{1.0}
\end{figure}

Rings viscously spread until the outer edge reaches the fluid Roche limit \citep{Murray99},
\be
	R_r \approx 2.46 R_p \lp\frac{\rho_p}{\rho_r} \rp^{1/3},
    \label{eq:rd}
\ee
where $\rho_p$ is the bulk density of the planet and $\rho_r$ is the density of particles that make up the rings. Outside of this radius, material aggregates into satellites and is no longer part of the ring. {At most, the rings can cover an area $f\pi R_r^2$, where $f\lesssim1$ is the filling factor that accounts for gaps and rings with optical depth less than unity (as is well-known for Saturn). The observed inferred radii then obey \mbox{$R_{\rm inf}\lesssim f^{1/2}R_r$,} because the exoplanet may be viewed at an oblique angle. Combining this with Equation (\ref{eq:rd}), we put a limit on the ring density,}
\be
    \rho_r &\lesssim& \lp\frac{2.46f^{1/2}}{R_{\rm inf}}\rp^3 \frac{3M_p}{4\pi}
    \nonumber
    \\
    &\approx& 2.3 f^{3/2}\lp\frac{M_p}{6\,M_\Earth} \rp
    \lp\frac{R_{\rm inf}}{6\,R_\Earth} \rp^{-3} {\rm g\,cm^{-3}}.
    \label{eq:rhor}
\ee
{Material with a density above this will not be able to make sufficiently large rings because it combines into satellites at large radii instead.} In Figure~\ref{fig:composition}, we plot this density for each of the super-puffs. We include error bars, which correspond to the current uncertainties in the mass and inferred radius for each planet, {and take $f=1$ since even smaller $\rho_r$ values are required for $f<1$.} The dashed horizontal lines roughly delineate the density range expected for rocky material. From this comparison, we see that Kepler~18d, 87c, 223d, and 223e could all be explained by rocky rings, while Kepler~177c is borderline. On the other hand, Kepler~51b, 51c, 51d, and 79d are so large that it is difficult to explain them with rocky rings unless the material is very porous \citep[although not out of the question, since some asteroids have densities as low as $\sim1.5\,{\rm g\,cm^{-3}}$;][]{Carry12}. It has been argued that the locations of the solar system rings might indicate that weak material that can easily be disrupted is required for generating rings \citep{Hedmam15}. Although speculative, this may explain the low densities inferred here for the super-puff ring material.

Another constraint is that the rings will be subject to Poynting-Robertson drag because of the relative close proximity to their parent stars \citep{Goldreich78}. Since the rings need to be optically thick to produce the large observed transits, the corresponding Poynting-Robertson timescale depends on the mass surface density $\Sigma$ \citep[rather than the particle size as is the case for typical Poynting-Robertson drag,][]{Schlichting11}, resulting in
\be
	t_{\rm PR} \approx \frac{\pi \Sigma c^2}{\sin i(5+\cos^2i)} \frac{4\pi a^2}{L_*},
\ee
where $i$ is the inclination of the ring with respect to the orbital plane and $L_*$ is the luminosity of the parent star. The total ring mass is roughly $M_r\approx 3\pi f\Sigma R_r^2/4$ \citep{Piro18b}. Using $R_{\rm inf}\lesssim R_r$, we can at least get a limit on the ring mass needed if the rings are to last a time $t_{\rm PR}$,
\be
    M_r \approx
    \frac{3t_{\rm PR}}{4c^2} \frac{fL_*}{4\pi a^2}
    R_{\rm inf}^2\sin i(5+\cos^2i)
    \label{eq:mr}
\ee
This mass is plotted for each of the super-puffs in Figure~\ref{fig:composition} for a timescale $t_{\rm PR}=10^9\,{\rm yr}$ and inclination of $i=45^\circ$ {(again with $f=1$, which gives an upper limit).} The range of masses is similar to large asteroid masses in our solar system \citep{Lang92}, showing that these are not unreasonable amounts of material for rings.

We note that another important timescale to consider is the viscous time for the rings, which should be dominated by collisions including self-gravity effects \citep[see discussions in][]{Daisaka01,Piro18b}. In detail, this depends on the exact density and size of the ring particles, but can easily be in the range of $\sim10^8-10^9\,{\rm yrs}$.

\subsection{Constraints on the Planetary Quadrupole Moment}

The presence of rocky rings with the desired properties also provides constraints for the underlying planets, which we explore in more detail next.

Another issue is that the rings must be oriented at an oblique angle with respect to the planet's orbital plane to produce the large transits. The ring's orientation depends on the competing effects of the planet's oblateness, quantified by the quadrupole gravitational harmonic $J_2$, and the tide from the parent star \citep{Tremaine09}. Equating these two effects provides an estimation of the so-called Laplace radius \citep{Schlichting11},
\be
    R_{\rm L}^5 \approx 2J_2M_pR_p^2\frac{a^3}{M_*},
\ee
where we have assumed that the orbital eccentricity is negligible \citep[justified by what is known for these multiplanet systems,][]{Fabrycky14,Hadden14,Hadden17}. Beyond this radius, a ring is warped into the orbital plane of the planet. Thus, we require $R_L\gtrsim R_r$ for the super-puffs. This implies a minimum $J_2$ of
\be
    J_2(R_{\rm L}=R_r) \approx \frac{(2.46)^2R_{\rm inf}^3}{2a^3}
    \lp\frac{M_*}{M_p} \rp\lp\frac{\rho_p}{\rho_r} \rp^{2/3},
    \label{eq:j2}
\ee
where we first use Equation (\ref{eq:rd}) to substitute $R_r$ for $R_p$ (since the radius of the underlying planet is unknown), and then we use $R_{\rm inf}\lesssim R_r$ to estimate $J_2$ from the current observables.

\begin{figure}
\epsscale{1.0}
  \includegraphics[width=0.48\textwidth,trim=0.2cm 0.0cm 0.0cm 0.0cm]{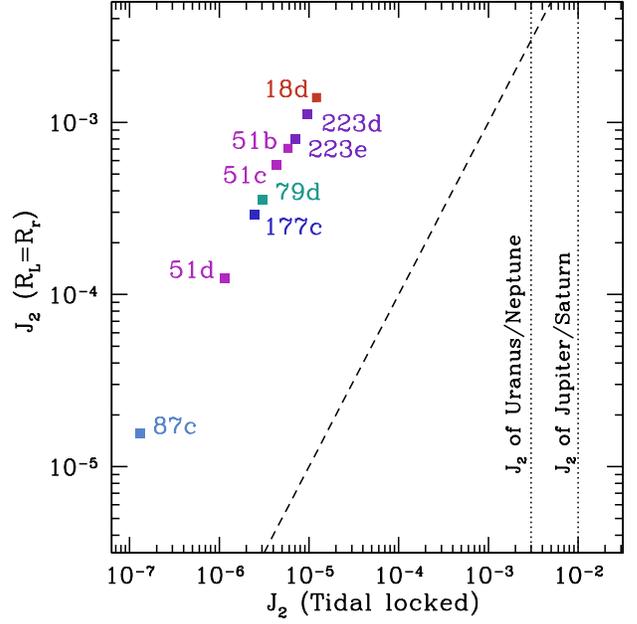}
\caption{A comparison of the two values for $J_2$ from Equations (\ref{eq:j2}) and (\ref{eq:j2 locked}). The diagonal dashed line shows where these two quantities are equal. Since all the super-puffs are to the left of this line, they are spinning too slowly and would have rings warped into their orbital planes if they are tidally locked. On the other hand, if the super-puffs have spins similar to the gas or ice giants in our solar system, and if they are able to prevent tidal locking, then their rings would not be warped.}
\label{fig:j2}
\epsscale{1.0}
\end{figure}

The actual $J_2$ of a super-puff depends on its rotation rate $\Omega$. This can be estimated as \citep{Chandrasekhar69}
\be
    J_2 \approx \Lambda\frac{\Omega^2R_p^3}{GM_p},
\ee
where $\Lambda\approx0.2-0.5$ is a factor that depends on the density distribution of the planet. Given the close proximity of these planets to their parent stars, it is natural to assume they are tidally locked. This would result in
\be
    J_2({\rm Tidal\ locked})
    &\approx& \Lambda\lp\frac{M_*}{M_p}\rp
    \lp\frac{R_p}{a} \rp^3
    \nonumber
    \\
    &\approx& 0.02\lp\frac{\Lambda}{0.3}\rp\lp\frac{M_*}{M_p}\rp
    \lp\frac{R_{\rm inf}}{a} \rp^3 
    \lp\frac{\rho_r}{\rho_p} \rp,
    \nonumber
    \\
    \label{eq:j2 locked}
\ee
where we again use Equation (\ref{eq:rd}) and $R_{\rm inf}\lesssim R_r$ to write this in terms of $R_{\rm inf}$. A comparison of the two values for $J_2$ from Equations (\ref{eq:j2}) and (\ref{eq:j2 locked}) is plotted in Figure~\ref{fig:j2}. We use \mbox{$\rho_r/\rho_p\sim2$,} corresponding to a rocky ring composition. In all cases, the $J_2$ implied for tidal locking is less than the $J_2$ needed to prevent ring warping. Therefore, if the super-puffs are tidally locked, then none of them can have rings at the inclinations needed to produce the large inferred radii.

On the other hand, the current $J_2$ of the gas and ice giants in our solar system are much larger with values of $\sim0.003-0.01$ as indicated on Figure~\ref{fig:j2}. If the super-puffs could have similar $J_2$ values, and not be impacted too drastically by tidal locking, then they could still have sufficiently large $J_2$ to prevent their rings from being warped.

\begin{figure}
\epsscale{1.0}
  \includegraphics[width=0.48\textwidth,trim=0.2cm 0.0cm 0.0cm 0.0cm]{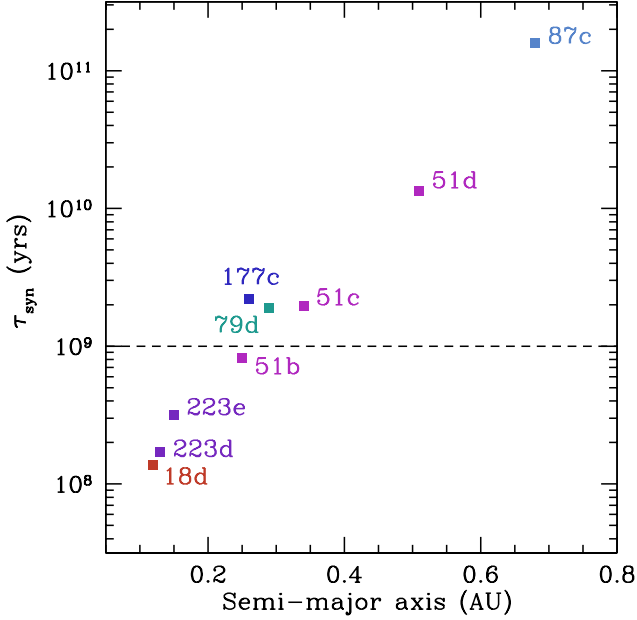}
\caption{Synchronization time $\tau_{\rm syn}$ for each of the super-puffs as a function of their semi-major axis $a$ for $Q_p=10^{6.5}$ and $k_2=3/2$. The dashed line delineates $10^9\,{\rm yrs}$. Planets above this line may be expected to maintain a $J_2$ independent of tidal locking.}
\label{fig:tau}
\epsscale{1.0}
\end{figure}

Motivated by this, we consider the tidal synchronization time for each super-puff, roughly estimated as \citep{Piro18a}
\be
    \tau_{\rm syn}
        \approx \frac{2\lambda Q_p}{3k_p}
            \lp\frac{M_p}{M_*} \rp
            \lp\frac{a}{R_p}\rp^3
            \frac{P}{2\pi},
\ee
where $\lambda\approx0.2-0.3$ is the radius of gyration, $Q_p$ is the tidal quality factor, and $k_p$ is the Love number. The resulting $\tau_{\rm syn}$ for each super-puff is summarized in Figure~\ref{fig:tau}, plotted using similar estimates as above for $R_p$, and  with $\lambda=0.3$, $Q_p=10^{6.5}$, and $k_p=3/2$. This shows that the tidal locking timescale is mostly a function of the semi-major axis, which is not surprising since $\tau_{\rm syn}\propto a^{9/2}$. The super-puffs with $\tau_{\rm syn}\gtrsim10^9\,{\rm yrs}$ may be able to maintain a sufficiently large $J_2$ to prevent ring warping. In contrast, Kepler~18d, 51b, 223d, and 223e may become tidally locked and have a smaller $J_2$. Even in these cases though $\tau_{\rm syn}\gtrsim10^9\,{\rm yrs}$, and these objects may still not be completely tidally locked if they are especially young \citep[such as for Kepler~51,][]{LibbyRoberts19} or if there is a factor of a few underestimate of the synchronization time. If the value of $Q_p$ was taken to be much smaller (for example, rocky planets like the Earth have $Q_p\sim10$), then all of the super-puffs would be tidally locked. Therefore, the super-puffs must have substantial gaseous envelopes even if they are explained by rings.

\begin{figure}
\epsscale{1.0}
  \includegraphics[width=0.47\textwidth,trim=0.0cm 0.0cm 0.5cm 2.5cm]{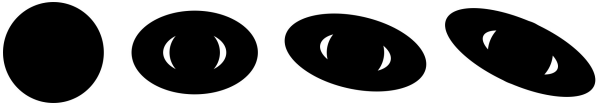}
\caption{{Example of a bare planet and three different ringed planets, each with the same covering area. From left to right, these are a bare planet with $R_p=6.98\,R_\Earth$, a ringed planet with $R_p=3.5\,R_\Earth$, $\theta=41.3^\circ$, $\phi=0$, a ringed planet with $R_p=4.0\,R_\Earth$, $\theta=33.0^\circ$, $\phi=22.5$, and a ringed planet with $R_p=4.5\,R_\Earth$, $\theta=34.1^\circ$, $\phi=45$.}}
\label{fig:18d_examples}
\epsscale{1.0}
\end{figure}

\begin{figure}
\epsscale{1.0}
  \includegraphics[width=0.45\textwidth,trim=0.0cm 0.0cm 0.0cm -1.0cm]{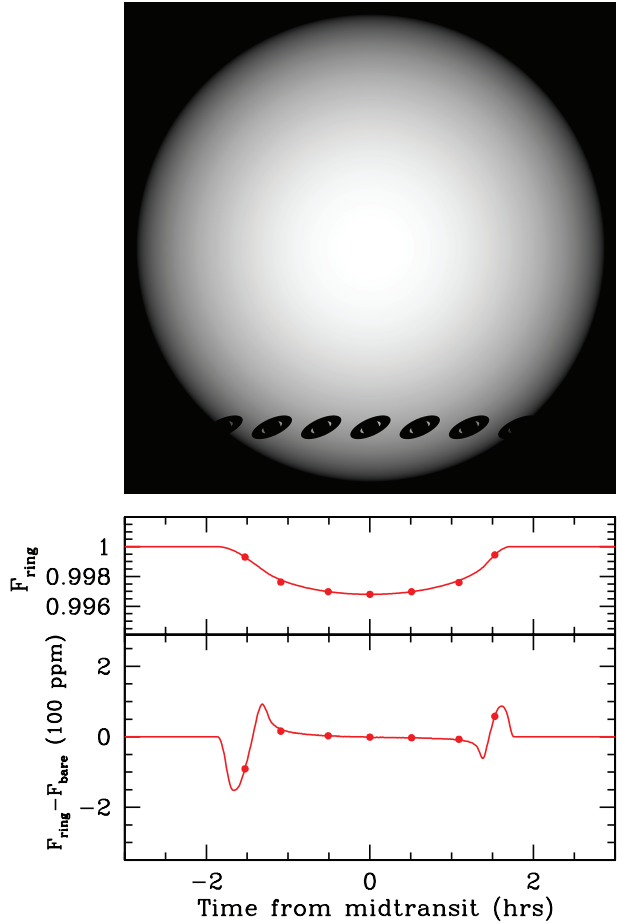}
\caption{Transit of a ringed planet and the resulting light curves. The upper panel shows an example of the images considered for the transit calculations with the planet at seven different positions. The middle panel shows the resulting transit (normalized to the stellar flux) with the points corresponding to each of the planet positions shown in the upper panel. The bottom shows the difference between a ringed transit and a bare star with the same covering area.}
\label{fig:multiplot}
\epsscale{1.0}
\end{figure}

\section{Detectability}
\label{sec:detectability}

If some of the super-puffs are actually ringed, then this can be revealed in the details of their transit light curves \citep[e.g.,][]{Barnes04,Santos15,Akinsanmi18}. Motivated by this, we simulate transits of ringed planets to assess whether this hypothesis is testable by current or future transit observations.

To construct the ringed planets, there are a few things to consider. First, the radius of the underlying planet $R_p$ must be chosen. This is unconstrained by the data, but must be in a range that gives a reasonable density given the planet's mass. Next, the inner and outer radius of the rings must be chosen. Since this depends on $\rho_p$ and $\rho_r$, both of which are unknown, we make the simple assumption that they extend from an inner radius of $\approx1.25R_p$ to $\approx2.5R_p$. Finally, we use the prescriptions summarized in \citet{Piro18b} to solve for the range of obliquities and ``seasons'' (the azimuthal angle of the planet) that together result in a transit depth that matches the observed $R_{\rm inf}$. {An example of a bare planet and three different ring sizes and orientations is shown in Figure \ref{fig:18d_examples}. This gives a sense for how a range of different silhouettes that can provide the same maximum transit.}

To perform these calculations, 5760 by 5760 pixel grayscale PNG images are generated for both the limb-darkened star and planet. For the limb-darkening prescription, we use the parameterization from \citet{Barnes03} with the coefficients $c_1=0.64$ and \mbox{$c_2=-0.065$.} These are simply chosen to mimic a realistic star for the examples presented here. In a true comparison to a specific super-puff, these coefficients should be fit for when modeling the transit. A useful feature of the super-puffs is that they are all in multi-transiting systems. Therefore the limb-darkening can be measured from the normal radius planets to be used for the super-puff ring fitting\footnote{As an aside, an initial assessment on whether some super-puffs have rings may be possible by looking for differences in the limb-darkening fit to individual planets orbiting the same star \citep{Akinsanmi18}}. We wrote a simple code to place the ringed planet at different locations across star, multiply the two images, and then sum up the pixels to find the total emitted light at any given time. These calculations do not include any forward scattering effects, since it has a relatively small impact in comparison to the many uncertainties for these systems. Since the strength of forward scattering depends on the size of the grains in the rings, its measurement may provide a more detailed understanding of the ring composition \citep{Barnes04}.
\begin{figure}
\epsscale{1.0}
  \includegraphics[width=0.48\textwidth,trim=0.0cm 0.0cm 0.0cm 0.0cm]{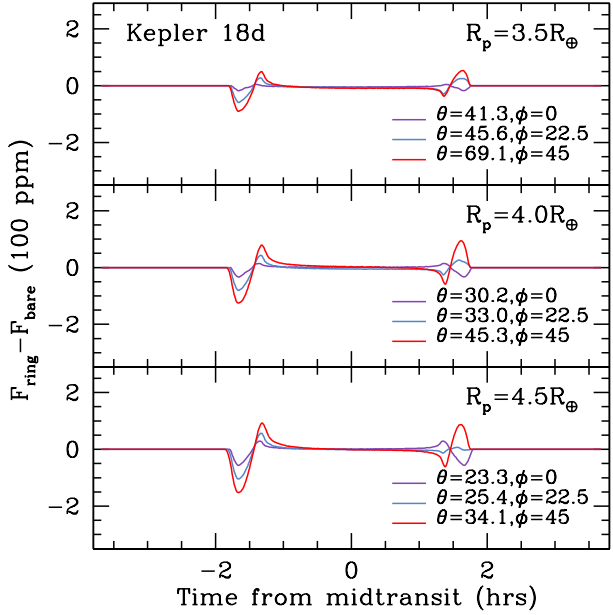}
\caption{A sampling of possible ringed exoplanet models that would provide the same transit depth as Kepler~18d. The three panels use planet radii of $3.5\,R_\Earth$, $4.0\,R_\Earth$, and $4.5\,R_\Earth$, from top to bottom. In each panel, three different combinations of angles are shown.}
\label{fig:18d}
\epsscale{1.0}
\end{figure}

One of these calculations is presented in Figure~\ref{fig:multiplot}. This example is a ringed planet with the same same projected surface area as Kepler~18b. The planet radius is assumed to be $R_p=4.5\,R_\Earth$ and the ring is positioned with obliquity $\theta=34.1^\circ$ and season $\phi=45^\circ$. The upper image shows the planet at seven different locations across the star's face. The middle and bottom panels shows the resulting transit and the difference between a ringed transit and a bare transit with the same surface area, respectively. Each point indicates a time of one of the snapshots from the top image.

This example demonstrates how the tilted projection of the ring naturally results in an asymmetric light curve. At ingress (on the left side) the ringed planet begins covering the star earlier than the bare planet would, resulting in a deficit of light. Conversely, at egress (the right side), the ringed planet stops covering the star because it is nearly parallel with the stellar limb, while a bare planet would still block light. This results in additional light at late times.

{In a more careful fit that compares bare and ringed transits, the stellar parameters should be fit in each case as well. This includes the stellar radius, limb darkening, and impact parameter of the planet. Here we use the values summarized in the literature from the fits assuming a bare planet for simplicity because our main goal is to qualitatively highlight how large the deviations are for ringed transits of super-puffs. A further complication when fitting super-puff transits is that they are all multi-planet systems, so the fits should account for both ringed and bare planets in the same systems to constrain the stellar parameters. Additionally, transit fits would also have to compare the evidence for a ringed planet model against that for an oblate planet model, as these two scenarios can produce similar transit shapes especially at high obliquity \citep{Barnes03, Akinsanmi18, Akinsanmi19} } 

{Beyond the complications with fitting ringed transits mentioned above, between the multitude of super-puffs, the range of underlying planet radii that each could have, the range of ring radii depending on the ring composition, and the many possible viewing angles, there are a multitude of possible parameters to consider for ringed planet transits.} For this reason, we focus on a few of the super-puffs as example cases and describe the general trends we have found.

\subsection{Kepler~18d}

The first super-puff we consider is Kepler~18d in Figure~\ref{fig:18d} (also highlighted previously in Figure~\ref{fig:multiplot}). This is an example of a relatively high density super-puff that is also close to its host star. For these reasons, Kepler~18d is attractive for having rocky rings because of the density required for its ring material, but less attractive because its relatively low synchronization timescale may mean that it is tidally locked to its host star and thus has a $J_2$ that is too low.

Figure~\ref{fig:18d} shows the results of our parameter survey for Kepler~18d. We consider radii of $3.5\,R_\Earth$, $4.0\,R_\Earth$, and $4.5\,R_\Earth$ for the underlying planet in the top, middle, and bottom panels, respectively. {These correspond to mean densities for the planet of $2.1$, $1.4$, and $1.0\,{\rm g\,cm^{-3}}$, respectively, which are in the range expected for Saturn- or Neptune-like planets.} We then construct rings with viewing angle required to match the $R_{\rm inf}$ measured for this super-puff. Three different seasons (values for the azimuthal angle $\phi$) are shown in each panel.

Some general trends can be seen across the nine models shown in Figure~\ref{fig:18d}. As we vary the season, the silhouette of the ringed planet becomes more asymmetric, and the differences between the ringed and bare transits becomes correspondingly more asymmetric as well. Furthermore, the differences become larger when the underlying planet has a larger radius. This may seem paradoxical, but the reason is that when the planet is larger, the ring is viewed more edge on (as seen by the angle $\theta$). This makes the ring silhouette less like a large planet.

\begin{figure}
\epsscale{1.0}
  \includegraphics[width=0.48\textwidth,trim=0.0cm 0.0cm 0.0cm 0.0cm]{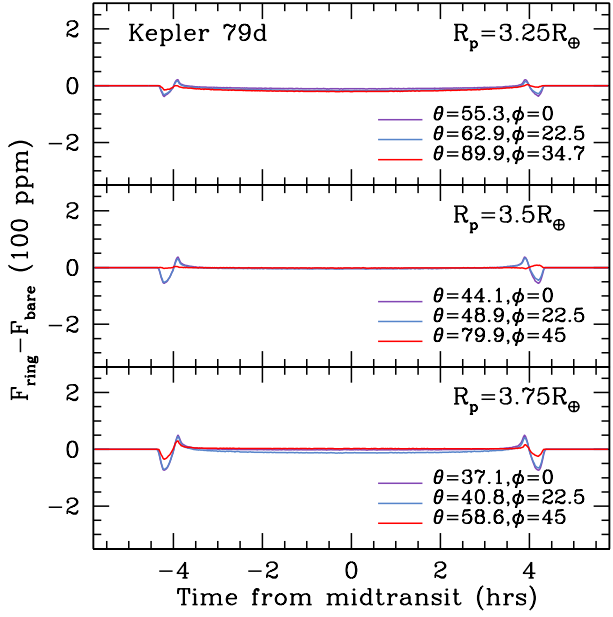}
\caption{The same as for Figure~\ref{fig:18d}, but for Kepler~79d.}
\label{fig:79d}
\epsscale{1.0}
\end{figure}

\begin{figure}
\epsscale{1.0}
  \includegraphics[width=0.48\textwidth,trim=0.0cm 0.0cm 0.0cm 0.0cm]{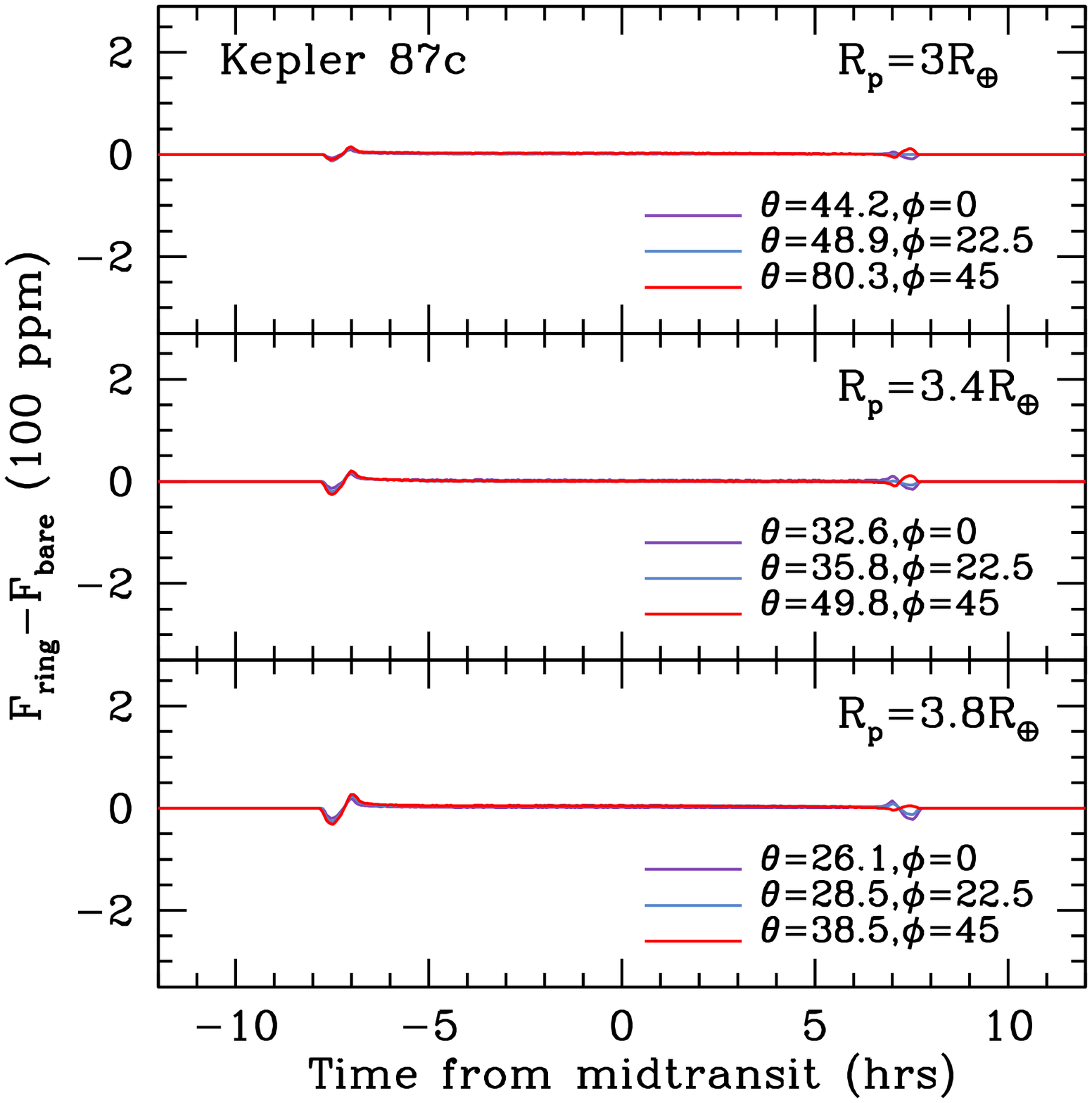}
\caption{The same as for Figure~\ref{fig:18d}, but for Kepler~87c.}
\label{fig:87c}
\epsscale{1.0}
\end{figure}

\begin{figure}
\epsscale{1.0}
  \includegraphics[width=0.48\textwidth,trim=0.0cm 0.0cm 0.0cm 0.0cm]{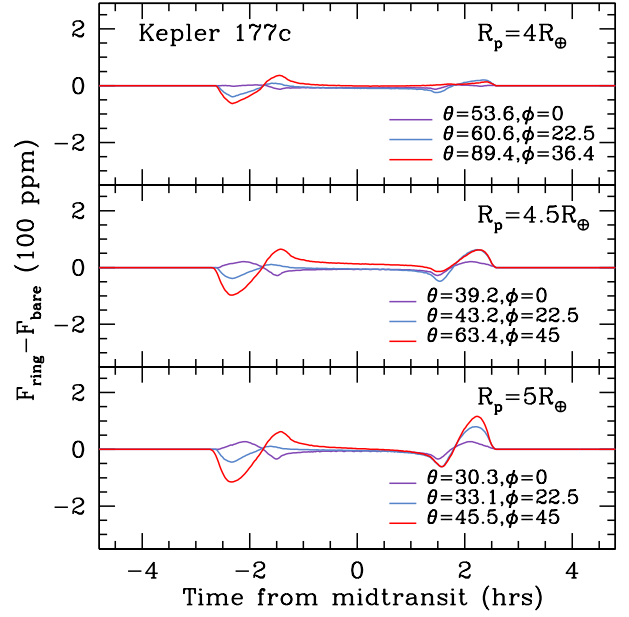}
\caption{The same as for Figure~\ref{fig:18d}, but for Kepler~177c.}
\label{fig:177c}
\epsscale{1.0}
\end{figure}

\subsection{Kepler~79d}

Next, we consider Kepler~79d as an example of an especially low density super-puff. The results are plotted in Figure~\ref{fig:79d}. {We again consider three different planet radii, but the range of potential radii are $3.25\,R_\Earth$, $3.5\,R_\Earth$, and $3.75\,R_\Earth$, which corresponds to mean densities for the planet of $0.96$, $0.77$, and $0.67\,{\rm g\,cm^{-3}}$, respectively. This range is smaller because it is more difficult to match the observed especially large radius inferred for this exoplanet, event with rings.} Furthermore, the obliquity of the planet must be relatively larger. This is because the ring must be viewed fairly face on to cover enough area. Correspondingly, the transit is more like an oblate planet rather than a ring and thus the difference between the ring and bare cases is at most $\sim50\,{\rm ppm}$. This is smaller than it was for Kepler~18d, where the residual signal could exceed $\sim150\,{\rm ppm}$.

\subsection{Kepler~87c and 177c}

{Finally, we consider the cases of Kepler~87c and 177c. As discussed in Section \ref{sec:basic}, a combination of factors for these super-puffs make them some of the most attractive for being explained by rings: the required $\rho_r$ is in the range for rocky material, the $J_2$ required for $R_{\rm L}\gtrsim R_r$ is small so that the rings would likely not be warped, and the long synchronization times means that they can maintain this $J_2$.

The results for Kepler~87c and 177c are summarized in Figures~\ref{fig:87c} and \ref{fig:177c}, respectively. For Kepler~87c, we consider planetary radii of $3\,R_\Earth$, $3.4\,R_\Earth$, and $3.8\,R_\Earth$, which correspond to mean planet densities of $1.3$, $0.90$, and $0.64\,{\rm g\,cm^{-3}}$, respectively. The variations are found to be small for Kepler~87c because of the relatively small size of this planet with respect to its parent star, as shown by  in Table~\ref{table}. Nevertheless, in the future photometry at a level of $\sim10-30\,{\rm ppm}$ should be able to determine whether the ring hypothesis works or not for Kepler~87c. For Kepler~177c, we consider radii of $4\,R_\Earth$, $4.5\,R_\Earth$, and $5\,R_\Earth$, which correspond to mean planet densities of $1.3$, $0.89$, and $0.65\,{\rm g\,cm^{-3}}$, respectively. Kepler~177c is more promising because the larger value of $(R_{\rm inf}/R_*)^2$ makes the photometric variations a factor of $\sim4$ larger.}

\subsection{Practical Considerations}

For the super-puffs that are most favorable for detecting rings (e.g., Kepler~18d and 177c above), measuring the residual signal from a ring requires $\sim$100~ppm photometry on $\sim$10~minute timescales, with shorter-period planets requiring finer temporal sampling. Diffuser-assisted photometry on ground-based telescopes has recently been able to approach this level of precision on brighter targets \citep{Stefansson17, Stefansson18, vonEssen19}. Considering the faintness of the super-puff host stars, however, this is a difficult goal. The $J$ magnitudes of the \textit{Kepler} planets in the super-puff sample range from $12-14$, for which the best precision on the necessary timescales is closer to $1000$~ppm for diffuser-assisted observations on $3-5$\,m telescopes \citep{Stefansson17,Vissapragada19}. Moving to 10\,m class ground-based facilities would improve the limiting precision, but not by an order of magnitude. {Stacking multiple transit observations may also improve the limiting precision, as the ring signal would be effectively static over multiple observations due to the long precession period. However, the difficulty of scheduling many transit observations for planets with long orbital periods and transit durations would realistically necessitate multi-year observing campaigns to build up the requisite baseline, even for the most favorable targets.} Observations aimed at detecting rings can thus only be performed with space-based facilities, at least for the \textit{Kepler} planets in the sample.

Searches for rings in the \textit{Kepler} sample have been attempted \citep{Aizawa18}. Although this work included the super-puffs Kepler~18d, 51b, 51d, and 79d, this analysis could only conclude that rings are not necessary to fit the currently available data. The \textit{Transiting Exoplanet Survey Satellite} \citep[\textit{TESS},][]{Ricker15} is not optimized to study these faint, long-period systems either. Using the Web TESS Viewing Tool\footnote{https://heasarc.gsfc.nasa.gov/cgi-bin/tess/webtess/wtv.py}, the expected 1~hr photometric precision for Kepler~18 is 1020~ppm; this does not take into account the difficulty of actually observing the transits due to the month-long \textit{TESS} Sectors.

In the case of \textit{Hubble Space Telescope} (\textit{HST}), identifying rings signals via space-based photometry is a similar problem to searching for exomoons, and \textit{HST} observations have provided evidence for an exomoon candidate around Kepler~1625b \citep{Teachey18}. Such searches for low-amplitude photometric signals can be compromised by time-correlated systematics, whether instrumental (e.g., \textit{Kepler}'s sudden pixel sensitivity dropout) or astrophysical (e.g., stellar variability) in nature \citep{Jenkins10, Christiansen13, Kipping12, Kipping15}. Additionally, different detrending methodologies seem to deviate at the required 100~ppm level even for \textit{HST} measurements of Kepler~1625, a star with similar brightness to the super-puffs studied here \citep{Teachey18, Kreidberg19, Teachey19, Heller19}.

Thus for secure photometry at the required precision, we must wait for the superior photometry of the \textit{James Webb Space Telescope} (\textit{JWST}), even for the most optimistic ring scenarios considered above \citep{Beichman14}. This adds to the list of predicted explanations for super-puff radii that are testable with \textit{JWST}; observations with \textit{JWST} may also be able to detect mid-infrared molecular features above a photochemical haze \citep{Kawashima19}, as well as diagnostic photometric features of a dusty outflow \citep{Wang19, LibbyRoberts19}.

\section{Conclusions and Discussion}
\label{sec:conclusions}

In this work we considered whether the super-puffs, planets with seemingly large radii for their masses, can be explained as ringed. We find that this hypothesis works better for some of the super-puffs and worse for others. Our main conclusions are as follows.
\begin{itemize}
\item The requirement that the rings be composed of rocky material favors Kepler~18d, 87c, 223d, and 223e as possibly being ringed. Kepler~177c is borderline.
\item The planets must be sufficiently oblate to prevent warping of the rings. This favors Kepler~51c, 51d, 79d, 87c, and 177c, both because of their long synchronization times and the low $J_2$ required to prevent warping.
\item Even if rings are present, the planets underlying super-puffs must still have substantial gaseous envelopes. {This is supported both by the densities we find for the underlying planets in our simulations and also to make their tidal locking timescales sufficiently long that $J_2$ can potentially be large.}
\item Taken together, rings likely cannot explain the entire super-puff population, but Kepler~87c and 177c have the best chance of being explained by rings. Kepler~18d, 223d, and 223e may also be interesting in case they are spinning faster than what is estimated here. Finally, Kepler~79d can only have rings if the ring material is especially porous.
\item Detection of rings via transits will be easiest to test for the higher density ($\gtrsim0.2\,{\rm g\,cm^{-3}}$) super-puffs or ones that have a higher overall signal as seen through the ratio $(R_{\rm inf}/R_p)^2$. This favors Kepler~18d, 51b, and 177c for testing this hypothesis.
\item Detection of rings will be hardest for planets that have the lower densities \mbox{($\lesssim0.1\,{\rm g\,cm^{-3}}$)} and smaller $(R_{\rm inf}/R_p)^2$, such as Kepler~79d, 87c, 223d, and 223e.
\item Except for HIP~41378f, which is discussed below, current ground- and space-based facilities are not precise enough to test the ring hypothesis. For the \textit{Kepler} super-puffs, such a test must wait for the launch of \textit{JWST}.
\end{itemize}
Confirmation of the presence of rocky rings in some cases would not only be an amazing new discovery, but also provide important information about these planets. This would allow a constraint on the obliquity and the spin through the quadrupole. Both of these would have implications for how the planets migrated to their current location, since even for the ringed hypothesis, it is likely that these planets were formed at larger stellocentric radii and migrated inward \citep{Lee16}. For rings to provide the necessarily large transits, the planets must maintain a large obliquity through this migration. Recent work shows that this may naturally be expected during the migration of closely-packed systems \citep{Millholland19a,Millholland19}, and the detection of rings would allow this affect to be directly measured.

During the finishing stages of this manuscript, HIP~41378f was announced as a new super-puff \citep{Santerne19}, which deserves some mention here. Super-puffs have primarily been characterized by transit-timing variations \citep[TTVs;][]{Holman05, Agol05, Agol18,Jontof-Hutter19}. For long-period planets, TTVs typically have more success than radial velocity (RV) measurements in identifying low-density planets due to a detection-sensitivity bias effect \citep{Mills17} as well as the intrinsic faintness of most super-puff systems. The measurement of the mass of HIP~41378f via RVs demonstrates that both techniques are capable of exploring super-puffs. This planet is especially exciting concerning the arguments presented here. Its large semi-major axis ($1.37\,{\rm AU}$) makes it less susceptible to having its rings warped, and for the bright parent star ($J = 7.98$), 100~ppm photometry is conceivable with current ground- and space-based facilities. The long transit duration and orbital period of HIP~41378f, however, makes it a difficult target to schedule for the required phase coverage \citep[global networks of high-precision photometers may alleviate this issue; see][]{vonEssen18}.

\acknowledgments
We thank Babatunde Akinsanmi, Jason Barnes, Konstantin Batygin, Eve Lee, Heather Knutson, Jessica Libby-Roberts, Alex Teachey, and Johanna Teske for helpful discussions. SV is supported by an NSF Graduate Research Fellowship and the Paul \& Daisy Soros Fellowship for New Americans.

\bibliographystyle{yahapj}

\end{document}